\documentclass[12pt,preprint]{aastex}

\slugcomment{}

\shorttitle{HUDF Star Formation}
\shortauthors{Thompson et al.}

\begin{document}

\title{Star Formation History of the Hubble Ultra Deep Field: \\
    Comparison with the HDFN}

\author{Rodger I. Thompson}
\affil{Steward Observatory, University of Arizona,
    Tucson, AZ 85721}
\email{rthompson@as.arizona.edu}

\author{Daniel Eisenstein}
\affil{Steward Observatory, University of Arizona,
    Tucson, AZ 85721}
\email{deisenstein@as.arizona.edu}

\author{Xiaohui Fan}
\affil{Steward Observatory, University of Arizona,
    Tucson, AZ 85721}
\email{fan@as.arizona.edu}

\author{Mark Dickinson}
\affil{National Optical Astronomy Observatories, Tucson, AZ 85719}
\email{med@noao.edu}

\author{Garth Illingworth}
\affil{University of California Santa Cruz, Santa Cruz, CA 95064}

\and

\author{Robert C. Kennicutt}
\affil{Institute of Astronomy, University of Cambridge, Cambridge CB3 OHA UK and Steward Observatory, 
	University of Arizona, Tucson, AZ 85721}
\email{robk@ast.cam.ac.uk}

\begin{abstract}

We use the NICMOS Treasury and ACS HUDF images to measure the extinction corrected star formation
history for 4681 galaxies in the region common to both images utilizing the 
star formation rate distribution function and other techniques
similar to those employed with the NICMOS and WFPC2 images in the HDFN.  Unlike 
the HDFN the NICMOS region of the HUDF appears to lack highly luminous and high 
star formation rate galaxies at redshifts beyond 3. The HUDF provides a region that 
is completely uncorrelated to the HDFN and therefore provides and independent measure 
of the star formation history of the universe.  The combined HUDF and HDFN star formation 
rates show an average star formation rate of 0.2 M$_{\sun}$ yr$^{-1}$.  The average SFR
of the combined fields at z = 1-3 is 0.29 M$_{\sun}$ yr$^{-1}$ while the average at
z = 4-6 is 1.2 M$_{\sun}$ yr$^{-1}$. The SFRs at all redshifts are within 3$\sigma$ of
the average over all redshifts.  

\end{abstract}


\section{Introduction}

The Hubble Deep and Ultra Deep Fields (\citet{wil96},\citet{thm99},\citet{dic00},
\citet{bec06}, \citet{thm05}) are our deepest look at individual objects in the 
early universe. There are numerous investigations of the properties of galaxies 
at high redshift utilizing the observations in the Hubble Deep Field North (HDFN)
(\citet{lan96}, \citet{mad96}, \citet{fer99}, \citet{thm01}, \citet{lan02}, \citet{thm03}
and others). The small size of the field, however, makes large scale structure the 
dominant error in projecting the properties of these galaxies such as 
the star formation history in the HDFN to the universe as a whole. The Hubble Ultra 
Deep Field (HUDF), with overlapping optical and near infrared images, provides an 
independent measure of the star formation history in a region completely uncorrelated 
to the HDFN.  As such it provides an independent very deep field data point for determining 
the history of star formation in the universe.  Unfortunately, even with a second
independent deep field, large scale structure is still a major error source in the 
quest for the history of the evolution of baryonic matter into stars and galaxies.  
These new observations do confirm, however, that the star formation rate (SFR) at
redshifts $\geq1$ is significantly higher than the present day rate determined by 
\citet{lil96}.  The results, however, are inconsistent with the star formation rates
that continue to increase to redshifts of 6 and beyond as found by \citet{lan02} in the HDFN.
Within the errors it is consistent with high redshift SFRs found in other areas 
by \citet{ste99}, \citet{bar00} and \citet{yan99}. It is also consistent with the 
SFRs found in the Northern and Southern GOODs field by \citet{gia04} which includes 
the HUDF.

The source extraction used in this investigation is relatively conservative and is
based on the combination of ACS and NICMOS fluxes.  For this reason we do not address
the issues of reionization, population III stars, and the luminosity function at
redshifts beyond 6.  These have been investigated with these same images by 
\citet{sta03}, \citet{sti04}, \citet{yan04} and \citet{bou04} with conflicting 
conclusions on the number, nature and evolution of high redshift sources. \citet{bou05}
has attempted to reconcile some of these conclusions.

\section{Observations}

\begin{figure}
\plotone{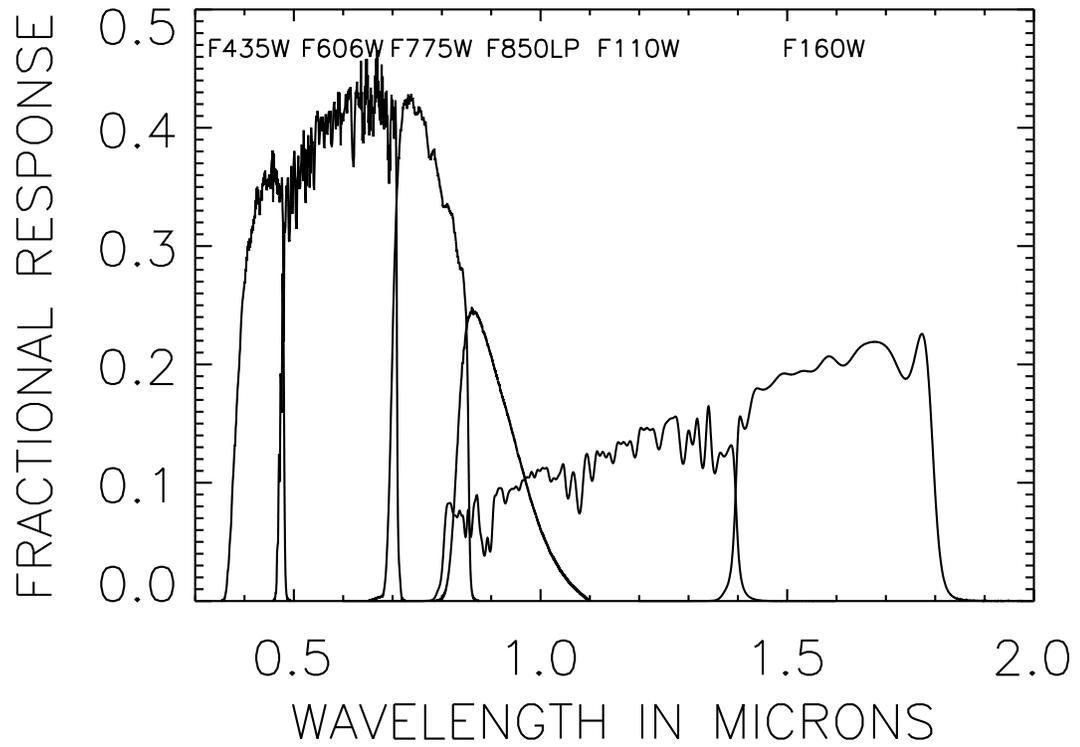}
\caption{The response functions for the ACS and NICMOS filters used
in the HUDF observations.  The response functions are the filter
transmission multiplied by the detector quantum efficiency at each
wavelength. \label{fig-fil}} 
\end{figure}

The observations are all taken in the HUDF centered at R.A.
$3\fh 32\fm 39.0\fs$  and Dec. $-27\arcdeg 47\arcmin 29.1\arcsec$ (J2000)  with the 
Advanced Camera for Surveys (ACS) and the Near Infrared Camera and Multi-Object
Spectrometer (NICMOS).  The ACS images were taken in the F450W, F606W,
F775W and F850LP filters and the NICMOS images were taken in the F110W
and F160W filters. The response curves for these filters is shown in
Figure~\ref{fig-fil}. The NICMOS images and data reduction are described
in detail in \citet{thm05} and the ACS images in \citet{bec06}.  The data
used in this analysis are only from the $144\arcsec$ by $144\arcsec$ region
where the NICMOS and ACS data overlap. In the rest of this manuscript we
will refer to the overlap area as the NUDF. The extent of the region was
determined by excluding pixels that had less than half the average 
integration time of pixels in the interior of the region.  This occurred 
due to the dithering of center points of the images on the 3x3 tile
pattern. Due to the significantly longer integration time per pixel, the 
ACS images have a higher signal to noise than the NICMOS images.  $30\%$
of the galaxies have F160W AB magnitudes of 28.0 or brighter which is the
approximate completness limit of the NICMOS images The depth of the 
NICMOS images are roughly equivalent to the 
depth in the HDFN \citep{thm03} while the ACS images are significantly deeper 
than the WFPC II images in the HDFN.  This should be remembered in the comparison 
of the results between the two fields.

The use of the ACS images in this study is not exactly equivalent to the
use of the WFPC2 images in the HDFN.  The ACS images lack the very blue 
F300W band produced by WFPC2 in the HDFN.  This makes the accuracy of low
redshift determinations less than is achievable with the F300W band. For 
very low redshift ($z \lesssim 1$) the F300W band provided a longer lever
arm by sampling the SED at shorter wavelengths and for somewhat higher 
redshifts ($ 2 \lesssim z \lesssim 3.5$) it
monitored the Lyman break at lower redshifts than the ACS F450W filter.
The ACS F850LP band also extends further to the red than the F814W band
and significantly overlaps the the NICMOS F110W band. The deep ACS images
in the F850LP provide high sensitivity to objects that would be dropouts
in the WFPC2 F814W filter but detected in the ACS F850LP filter.  The ACS 
images are better sampled than the WFPC2 images, although our rebinning of the ACS
image to match the NICMOS resolution negates most of that advantage.  The
spatial resolution of the NICMOS images is slightly less than was available
for the HDFN observations since the HST secondary was not refocused to achieve
the best NICMOS resolution as was done in the two HDFN NICMOS campaigns.  The
difference in resolution is only about $10\%$, however, and should not affect the
photometry of extended sources.

\section{Data Reduction} \label{s-dr}

The data reduction of the NICMOS bands is described in detail in \citet{thm05}
except for the source extraction procedure which is described in \S~\ref{ss-sz}.
The ACS images were taken directly from the HST Treasury Archive maintained
by MAST.  The $0.03\arcsec$ ACS pixels were binned to match the $0.09\arcsec$ drizzled 
NICMOS images by simple addition of the ACS images in 3 by 3 pixel bins.  The
NICMOS images were then aligned to the ACS images by a three step process.  The
initial alignment was done by using World Coordinate System values in the 
headers of the images.  The image alignment was then improved by cross correlating
the F110W NICMOS image with the F850LP image in 0.1 pixel steps.  The final alignment
was done by finding the centroids of compact objects and aligning the images by
the centroids.  In all cases the NICMOS images were determined relative to the ACS 
images. The image positions of each NICMOS image relative to the ACS image were
then used in the drizzle procedure to produce a mosaic NICMOS image that is
accurately aligned with the ACS F850LP image and hence with the ACS images
in all of its bands.
In each orbit the F160W image immediately followed the F110W image without
moving the spacecraft, therefore, the alignments calculated for the F110W images
were also applied to the F160W images. The 4 ACS and 2 NICMOS images used for
the source extraction are available from MAST in the NICMOS UDF Treasury 
version 2 files.  A mask was used to limit the extraction to only the area
that contained valid NICMOS images, an area of $144\arcsec$ on a side which is 
less than half of the area of the ACS UDF field. 

\subsection{Source Extraction} \label{ss-sz}

\begin{figure}
\plotone{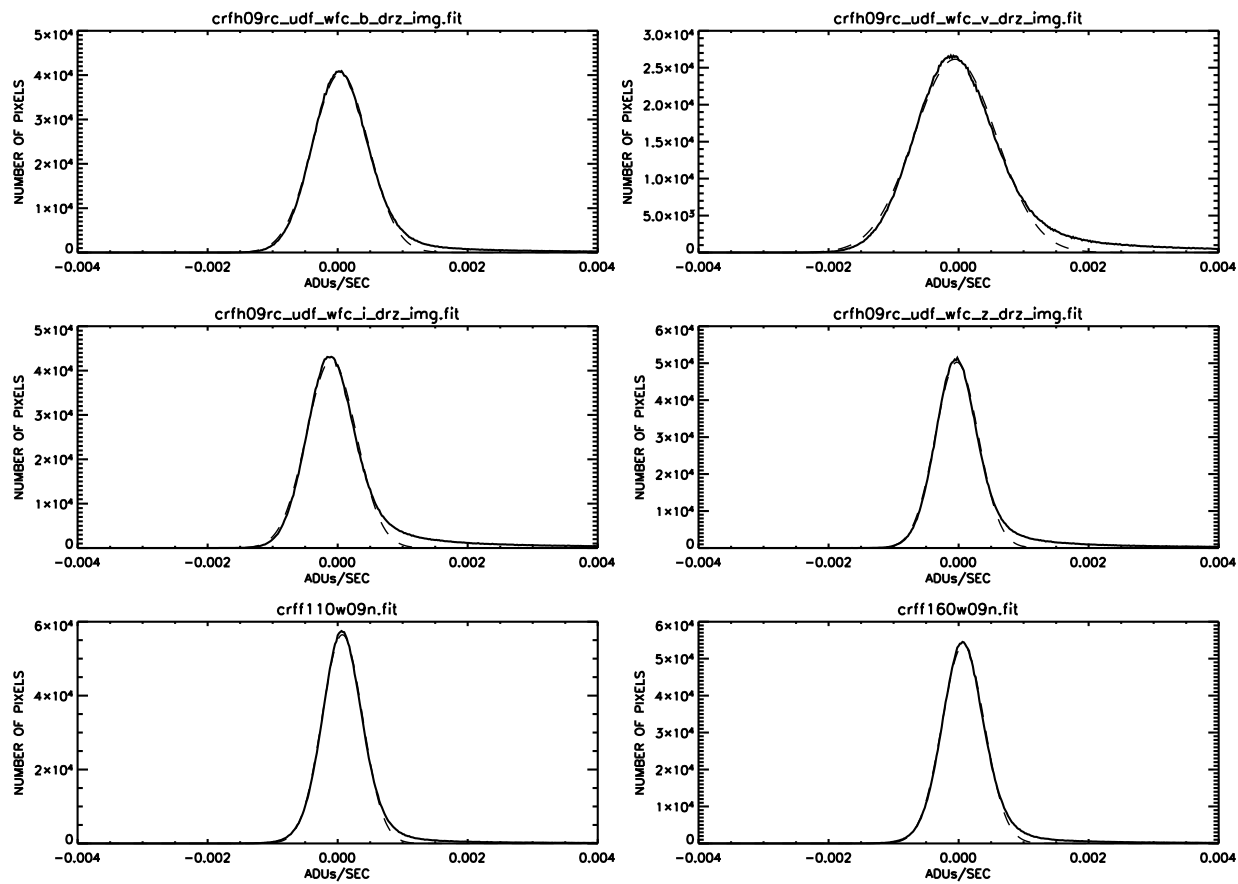}
\caption{Histograms of the distribution of pixel values in the six
images used in this study.  The signals are in adus per second.  The
dashed line is a Gaussian fit to the distribution.  The width of the
Gaussian is used to find the average standard deviation of the noise in the
image.  The positive side power above the Gaussian fits is due to
real sources in the image. \label{fig-sig}}
\end{figure}

Unlike the creation of the NUDF Treasury catalog, source extraction 
is done in two steps and is based on the fluxes in all ACS and NICMOS bands, 
not just the two NICMOS bands.  The first step is to identify
pixels that have sufficient signal to noise to be considered source
pixels and second to identify sources with the extraction program
SExtractor, hereinafter SE \citep{ber96}.  Source pixel identification
uses the procedure developed by \citet{sza99} for multiple band images.
The first step in the process is the production of images with unit
variances.  For the NICMOS images we first constructed inverse quantum 
efficiency (QE) maps from the flat field images for the F110W and
F160W camera 3 filters. The flat field images were drizzled in exactly
the same way as the F110W and F160W images.  Since the NICMOS flat
fields are multiplicative this procedure produced inverse QE maps
with the same weighting as the images. The F110W and F160W standard 
deviations from Figure~\ref{fig-sig} were multiplied by 1.8 to 
account for drizzle correlation \citep{fru02}.  The inverse QE maps were then 
multiplied by the standard deviations to produce a standard deviation
map.  The images were then divided by the standard deviation map
to produce an image with unit standard deviation and hence unit
variance.  The same procedure was carried out with the ACS images
except that only the drizzle weights were used rather than drizzled
flat fields due to the much more uniform QE of the ACS detectors.

\begin{figure}
\plotone{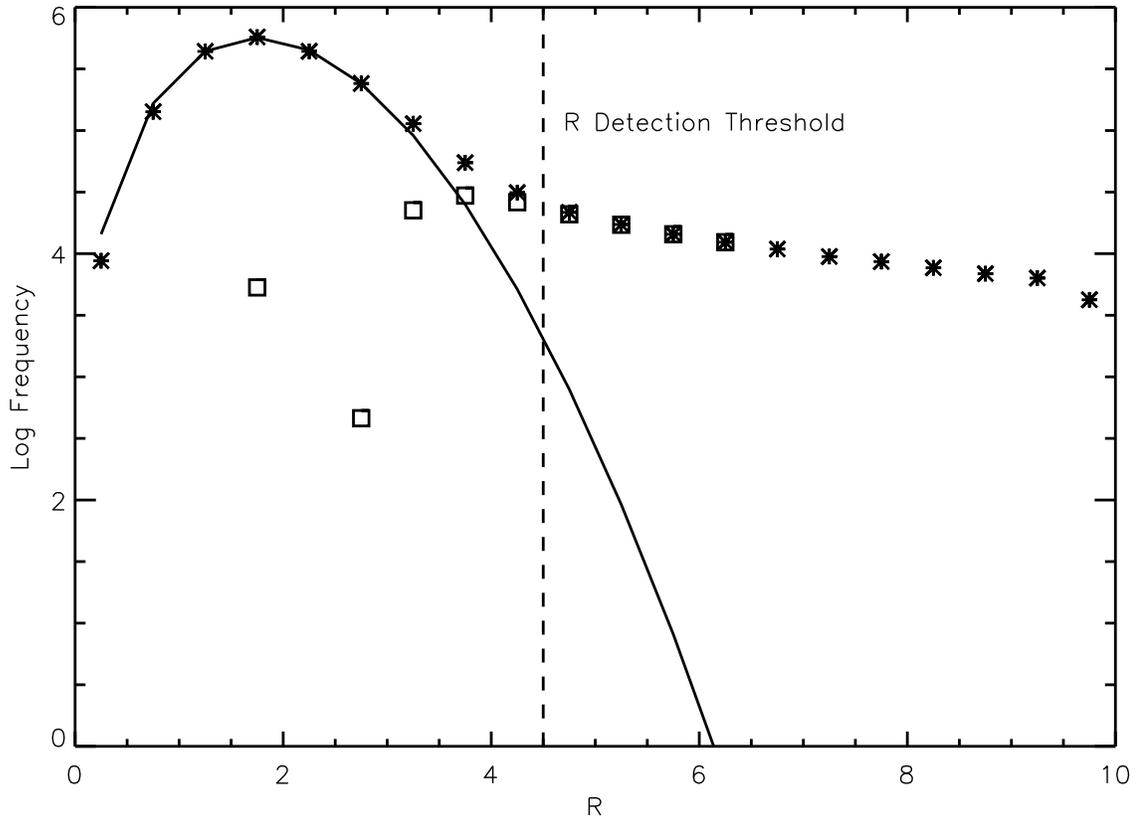}
\caption{Histogram of the Szalay et al. R values as described in
the text for the pixels in the NICMOS
region of the UDF.  The dashed line indicates the threshold R
value of 4.5 for source pixels.  The asterisks are the histogram
values, the solid line is the R probability distribution for 4
degrees of freedom and the squares are the histogram values minus
the 4 degree of freedom probability distribution.  See \S~\ref{ss-sz}
for a discussion of the method. \label{fig-szr}}
\end{figure}

Each pixel is assigned a $g_i$ value of 

\begin{equation}
g_i = \frac{f_i - s_i}{\sigma_i}
\end{equation}

\noindent where $f_i$ is the flux, $s_i$ is the sky background, and 
$\sigma_i$ is the standard deviation and $i$ refers to the six bands.  
In this case the sky background has been subtracted and the fluxes 
come from the images that have been scaled to have unit variance which
makes $\sigma$ values also unity.  This makes the $g_i$ values equal to  
the $f_i$ values.  Next each pixel is given a $y$ value which is

\begin{equation}
y = \sum_{i=1}^{6} g_i^2
\end{equation}

\noindent for every pixel.  The image formed by the y values is called the R
map which should be a map of the $\chi^2$ values with 6 degrees of freedom.
The histogram of the R values is shown in Figure~\ref{fig-szr} along
with the probability distribution for $\chi^2$ for \emph{4}
degrees of freedom which is a much better fit than the distribution for 6 
degrees of freedom.  This is most likely because the variance of the F110W
and F160W images has been overestimated and that the unit variance images for
these two bands are too low.  The NICMOS images have a non-Gaussian distribution
of noise values due to the correlation introduced by the drizzle procedure as
described by \citet{fru02}.  The tail is visible in Figure 3 of \citet{thm05}
on the negative side of the noise distribution.  Flux from sources obscures the
positive tail, however, it is present.  Scaling of the NICMOS images to higher
values reproduces the $6$ degree of freedom distribution but the non-Gaussian
tail produces spurious sources.  For this reason we chose not to scale up the 
NICMOS images.  This does not significantly affect the source extraction since 
the threshold R value is tuned to produce a thorough but conservative source 
extraction. The ACS images do not contain a non-Gaussian component since they 
were drizzled using a point target which does not spread flux between target 
pixels.  In addition they have been rebinned in boxes of 3 by 3 pixels which 
greatly reduces any cross pixel correlation.  This illustrates one of the difficulties
in dealing with two data sets with significantly different signal to noise 
ratios and different noise characteristics.
The boxes in Figure~\ref{fig-szr} show the observed histogram of R values represented
by the asterisks minus the noise value of R as determined by the probability
distribution traced by the solid line. The boxes represent the number of source 
pixels for a given R value.  The vertical dashed line at R = $4.5$ is the R 
threshold used in this work for source pixels.  

A new image is produced that has the R value multiplied by 1000 for all pixels
above the threshold and all other pixels at a random noise level with a standard 
deviation of $10^{-6}$.  Without the minimal random noise the next
step, source extraction with SE, aborts.  Sources are extracted with
SE in the two image method.  The first image is the image derived
from the R map and the second is the image in one of the ACS or NICMOS
bands.  Used in this way the sources are found from the first image
but the photometry is extracted from the second.  The configuration
files for SE use an absolute flux threshold which is set so that all
pixels identified by the Szalay et al. procedure are above the threshold and
all pixels in the random noise field are orders of magnitude below
the threshold.  In this procedure the only role that SE plays in the
source selection process is to enforce the criteria that any legitimate
source must have a minimum area of 4 pixels and to identify the sources
according to the deblending criteria.  The value of DEBLEND\_NTHRESH 
was set to 32 and the value of DEBLEND\_MINCONT was set to 0.0001.  The
extraction produced 4702 sources.  Two of these sources are bright stars
with clear diffraction spikes.  22 more unresolved objects were identified
from the list of \citet{pir05}.  These objects were not included in the 
determination of the star formation rates.  Inspection of all of the sources 
revealed 19 sources that were suspect due to noise in the F110W band. 
Noise in the F110W band in generally due to cosmic ray persistence from
hits that occurred before the beginning of an integration.  As described
in \citet{thm05}, the F110W band integration was always the first integration
after earth occultation and subject to persistent hits that occurred
during occultation.  Hits that occur during the F110W integration or the
immediately following F160W integration are removed by the cosmic ray hit
removal software.  The hits are recognizable by either linear features or
cosmic ray spray geometries. The 19 suspect objects and the two stars were removed
leaving 4681 sources. Removal of these sources had no affect on the conclusions
of this paper. Although most of the remaining sources have no detectable signal
in the two NICMOS bands the NICMOS upper bounds are useful in constraining the 
spectral energy distributions (SEDs).

The source extraction returns the flux in three apertures of diameters $0.54\arcsec$,
$1.08\arcsec$ and $3.06\arcsec$ along with the isophotal and SE defined best and
auto apertures.  SE also returns a detection image with all of the pixels
defined as part of an object labeled with the source ID.  It is important to
remember that in this case the pixels were not determined by SE but by the
Szalay procedure described above.  SE simply distributed the pixels into
individual objects.  In the following analysis we only use the aperture 
fluxes for determining the source properties and then sum over all pixels
of the object to determine total fluxes.

Inspection of the images and extracted objects clearly indicates that there 
are valid sources that have been missed by the extraction process.  For example
none of the z $\geq$ 7 sources described by \citet{bou04} which only have flux
in the NICMOS bands were extracted.
Reduction of the detection threshold below 4.5 yields more sources but also
starts to pick up sources that may be spurious.  In this analysis we decide
to take a conservative approach to source extraction and choose a threshold
that has a high confidence level.  A more aggressive source extraction, with
a reliable check on source confidence may yield additional information on
star formation at higher redshifts.  The corrections for missed luminosity
described in \S~\ref{ss-hx} will compensate in the SFR determination for objects 
missed due to our conservative threshold.

\section{Sensitivity and Aperture Correction}

The initial SED analysis produced a large number of sources with extremely
blue NICMOS F110W to ACS F850LP colors but with reasonable NICMOS F160W to
NICMOS F110W and ACS F850LP to ACS F775W colors.  This led to a reanalysis
of the NICMOS F110W and F160W sensitivities based on the large number of
NICMOS Camera 3 observations of the solar analog star P330-E which has 
served as a NICMOS calibration star based on ground based photometry by
\citet{per98}.  This analysis revealed that the two bands were approximately
$10\%$ less sensitive than the analysis at the beginning of the NCS era had
found which was based on relatively few observations.  At the same time 
detailed aperture corrections were also performed which led to increased
aperture corrections in going from the $0.54\arcsec$ diameter aperture measurements
to the full power measurement for a point source.  This new analysis is
described more fully in \citet{thm05b}. Due to the very high signal to noise ratio
of the P330 observations the aperture corrections are far more accurate than the
faint photometry and hence to not contribute to the noise budget of the observations.
The net effect of these changes
in calibration resulted in an increase in the NICMOS fluxes making the NICMOS 
to ACS colors consistent with the ACS to ACS colors and NICMOS to NICMOS
colors.  The fluxes in the Version 
2.0 Treasury Catalogs in the MultiMission Archives at STScI (MAST) have
not had the new sensitivities applied.  Also note that the aperture fluxes
and magnitudes in the catalog are for the photons detected in the stated
apertures and must be corrected for the aperture function given in Table~\ref{tab-ap} or
\citet{thm05b} when comparing with measurements with other instruments or
facilities such as ACS.  Table~\ref{tab-ap} lists the aperture corrections
used in this work for the $0.54\arcsec$ diameter aperture used to define the 
galaxy SED.  The ACS filter aperture corrections are taken from \citet{sir05}.

Although the aperture corrections are for point sources, they are
accurate for the small aperture ($0.54\arcsec$) used for the parameter analysis.
This small aperture flux is only used to determine the parameters of redshift,
extinction and SED as described in \S~\ref{ss-sp}.  All photometric parameters
are found by summing all of the pixels identified by SE as part of the object.
They also have the merit of being accurately determined by this study and 
\citet{sir05}.  As a test an alternative method was carried out in which a 
scaled NICMOS PSF, as determined from the P330E image described above, was 
used to convolve the ACS images so that the PSF of the ACS image reasonably
matched the NICMOS image.  The NICMOS PSF was scaled since a convolution
with the full width NICMOS PSF on the 3x3 binned ACS data would produce an
ACS PSF that was broader than the NICMOS PSF.  The full analysis from source
extraction through SFR determination was performed and compared to the analysis
using aperture corrections.  In every redshift bin other than z = 6 the SFR
was equal within $1\sigma$.  At z = 6 the SFRs were equal within $2\sigma$
with the SFR from the convolved images being lower than the aperture correction
SFR due to some galaxies having lower extinction and earlier SED values in
the convolved image reduction compared to the aperture correction reduction.
These changes were within the Poisson distribution of parameter errors as 
determined from the photometric error analysis described in \S\ref{ss-er}.
We did not discover any systematic trends in redshift, SED or extinction
between the two samples.

At this time there is some debate on whether NICMOS observations have a small
nonlinear response in the sense that observations of faint sources may have
lower sensitivity than brighter sources.  To measure the sensitivity of our
results to such an effect, the analysis was repeated using a NICMOS response
function in both filters that was $10\%$ less than the one determined above.
The net change in SFR in each redshift bin was less than the errors shown in 
Figure~\ref{fig-sfr}.  We therefore proceeded with the response functions 
determined from the observations of P330E.

\section{Data Analysis} \label{s-da}

After the extraction of the source photometry the redshift, extinction,
and SEDs of all of the sources are determined.  The three parameters
are determined using the $0.54\arcsec$ diameter aperture fluxes corrected
by the factors in Table~\ref{tab-ap}.  The $0.054\arcsec$ aperture fluxes
are the most appropriate flux measurement for parameter determination for
several reasons.  Larger aperture or isophotal fluxes often are contaminated
by flux from adjacent sources.  Also the aperture fluxes include the highest signal
to noise regions of the galaxy and experience with the HDFN has shown that
they are the least susceptible to parameter perturbation due to photometric
errors. 

\subsection{Source Parameters} \label{ss-sp}

As described in previous work (\citet{thm01},\citet{thm03}), the source
parameters of redshift, extinction and intrinsic SED are determined by
a $\chi$ squared analysis of a comparison of the observed flux in the
6 photometric bands to the fluxes predicted by a suite of SEDs that have
been numerically redshifted and extincted using the obscuration law of
\citet{cal00}. This analysis include both a determination of the 
extinction and an error term in the $\chi$\ squared analysis that is
proportional to the flux to account for errors in the absolute 
photometric calibration of ACS and NICMOS (see Eqn. 2 in \citet{thm03}).
In this analysis the percentage error is set to $5\%$ instead of $10\%$. 
This better reflects the expected error in the ACS images and the recalibrated
photometry match between NICMOS and ACS.

\subsubsection{Template SEDs} \label{sss-ts}

\begin{figure}
\plotone{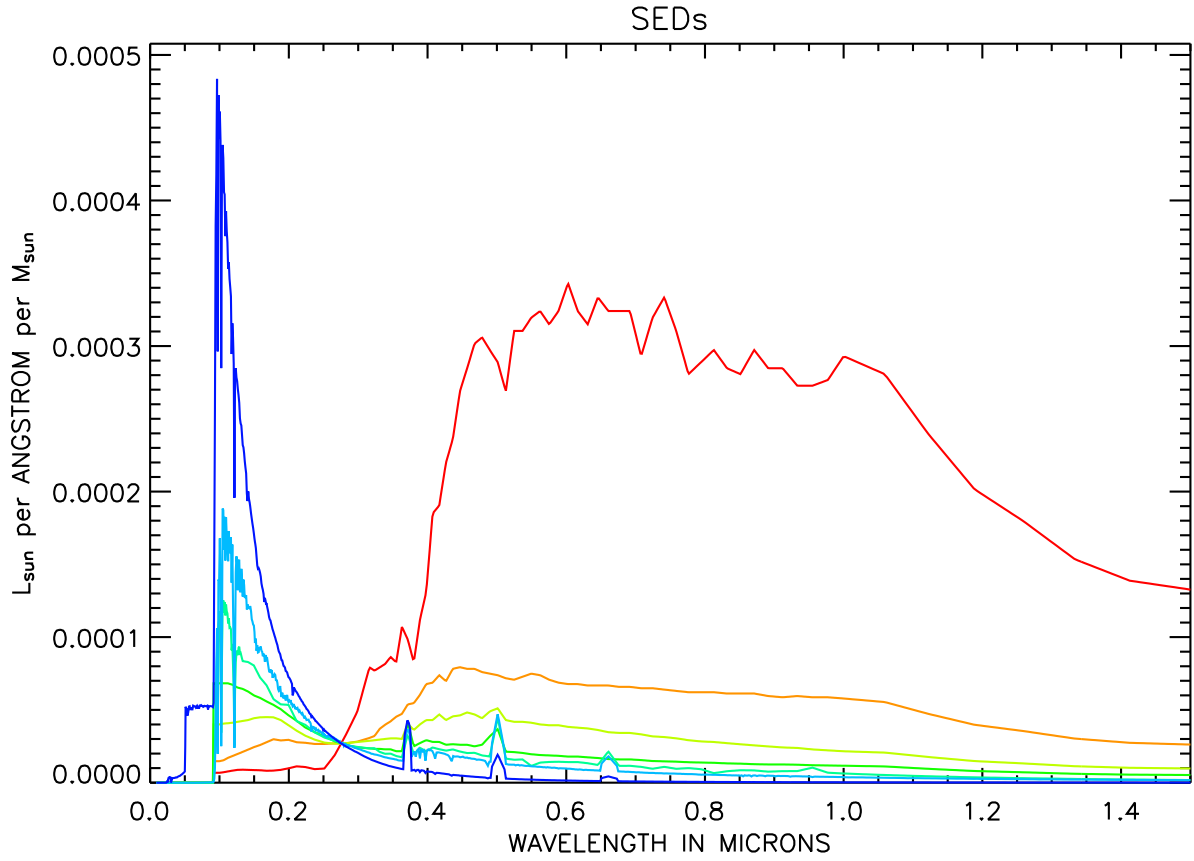}
\caption{The 7 template SEDs used in this analysis.  The two bluest
templates have been calculated from the models of \citet{bru03} as
described in the text. \label{fig-sed}}
\end{figure}

Two changes were made from the previous analyses.  A bluer and hotter
calculated SED template was added to make 7 basic templates rather
than the previous 6 templates.  This was done because many of the 
galaxies were bluer than the bluest previous template which was for
a 50 million year old galaxy.  The new galaxy SED is a template 
calculated for a 10 million year old galaxy with a metallicity of
0.0001 and a \citet{cha03} IMF from the SEDs of \citet{bru03} based
on the recommended Padova 1994 models.  Also the previous 50 million
year old template 6 model based on the BC96 models was replaced with
a 50 million year old \citet{bru03} model with a metallicity of 0.004
and a \citet{cha03} IMF.  These models have labels of m22 and m42
respectively in the \citet{bru03} library.  The first 5 templates
based on observed SEDs are unchanged from the previous analysis as
described in \citet{thm03}.  The SEDs are shown in Figure~\ref{fig-sed}.
Following past procedures the fluxes predicted by these base templates are
interpolated between the base SEDs to produce 9 intermediate templates
between each base template for a total of 61 total templates labeled
x.y where x is the base (1-7) and y is the increment (1-9). Note that
the first 4 observed templates from \citet{col80} were chosen by them
to have very low extinction and template 5 from \citet{cal94} is corrected
to zero extinction.  The small residual extinction in the observed galaxies
has no significant effect on the derived SFRs, particularly because the
vast majority of the star formation comes from galaxies with the latest
3 SEDs.

\subsubsection{Extinction} \label{sss-ex}

For this analysis we have switched to the Calzetti obscuration law
described by \citet{cal00}.  Otherwise the analysis is identical to
the previous analysis.  The E(B-V) values listed for the sources is
the E(B-V)$_s$ value described in \citet{cal00} which is the value
of E(B-V) for the stellar continuum rather than the E(B-V) for the
nebular continuum.

\subsubsection{Redshift values}

The allowed redshift values ranged from 0 to 10, however, only the 
redshift range up to 6.5 was used in the star formation rate analysis.
14 of 4681 galaxies have redshifts above 6.5.  Three galaxies have 
redshifts of 10.0, E(B-V) of 1.0, the maximum, and SED template values of 
2.0, the second earliest base template.  It is clear that these galaxies had 
properties that were not well matched by 
the templates and were driven to the
red boundaries of the extinction and redshift space.  Visual inspection
of the fluxes of these galaxies revealed that they are relatively faint 
galaxies with positive noise in the NICMOS F110W or F160W band which was 
interpreted as a high redshift Lyman break. This is one of the problems  
that occurs when two data sets with quite different signal to noise 
properties are mixed.  The problem was resolved by setting the F160W
flux to zero for these objects, forcing the fit to depend more heavily
on the optical bands.  This is not the most satisfying solution but
the contribution to the star formation rate of these objects is negligible.

\begin{figure}
\epsscale{.8}
\plotone{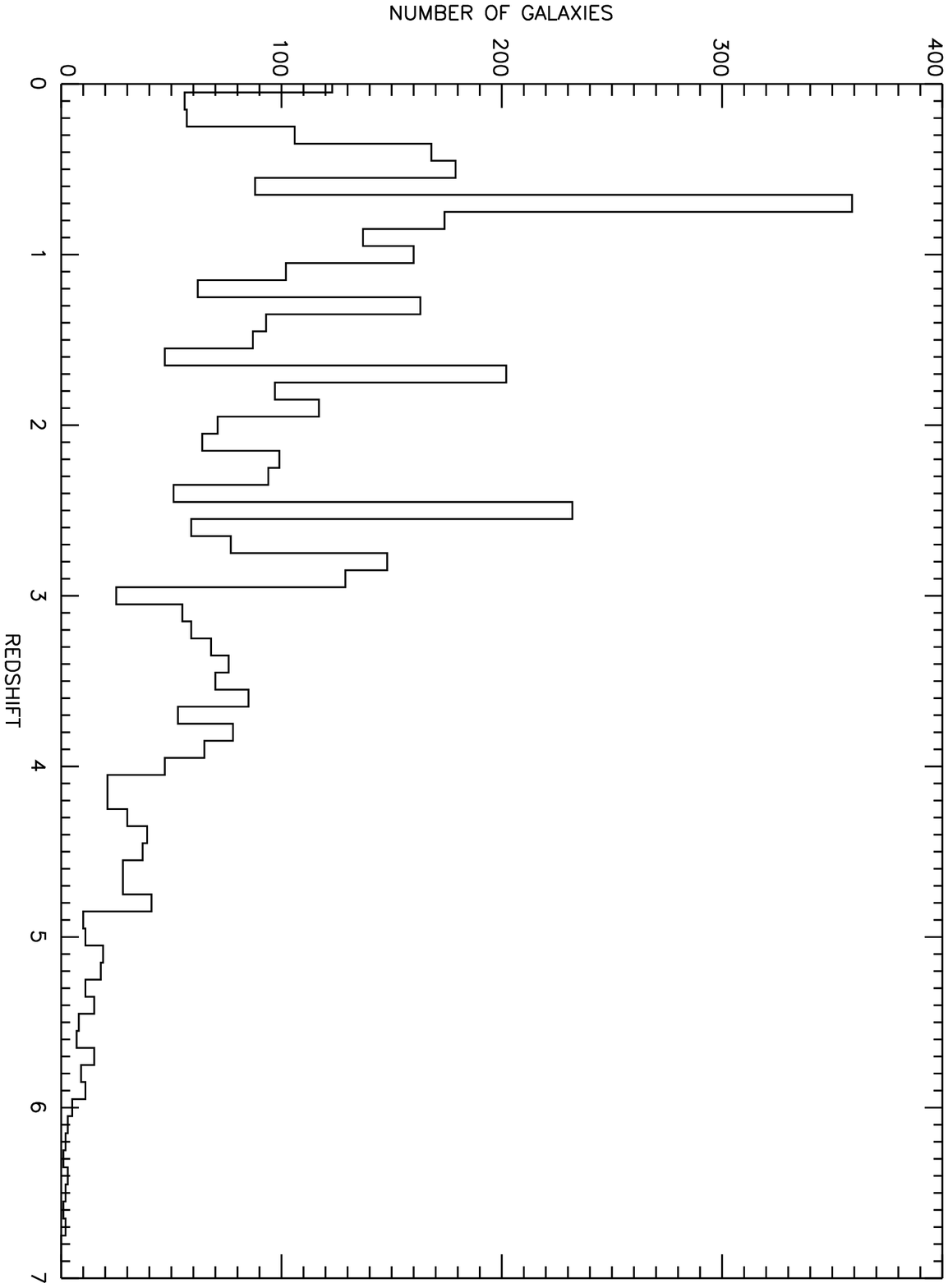}
\epsscale{1}
\caption{A histogram of the redshift values for the galaxies in the
NICMOS region of the UDF.  Only galaxies in the redshift region 
between 0.5 and 6.5 are used in this analysis. \label{fig-zh}}
\end{figure}

For 33 of the objects in the NUDF there are known redshifts from the
VIMOS survey in the Chandra Deep Field South (CDFS).  These are mostly at
low redshift ($z \leq 1$).  For these 33 objects the extinction and SED 
were chosen by finding the minimum $\chi$ squared value among the calculated
$\chi$ squared values in extinction and SED at the nearest grid redshift to
the known redshift. This is different than what was done in \citet{thm03} where 
the redshift was set to the known redshift after the $\chi$ squared analysis which 
did not necessarily lead to the best SED and extinction for the known redshift. The 
histogram of redshift values is given in Figure~\ref{fig-zh}.  

\begin{figure}
\plotone{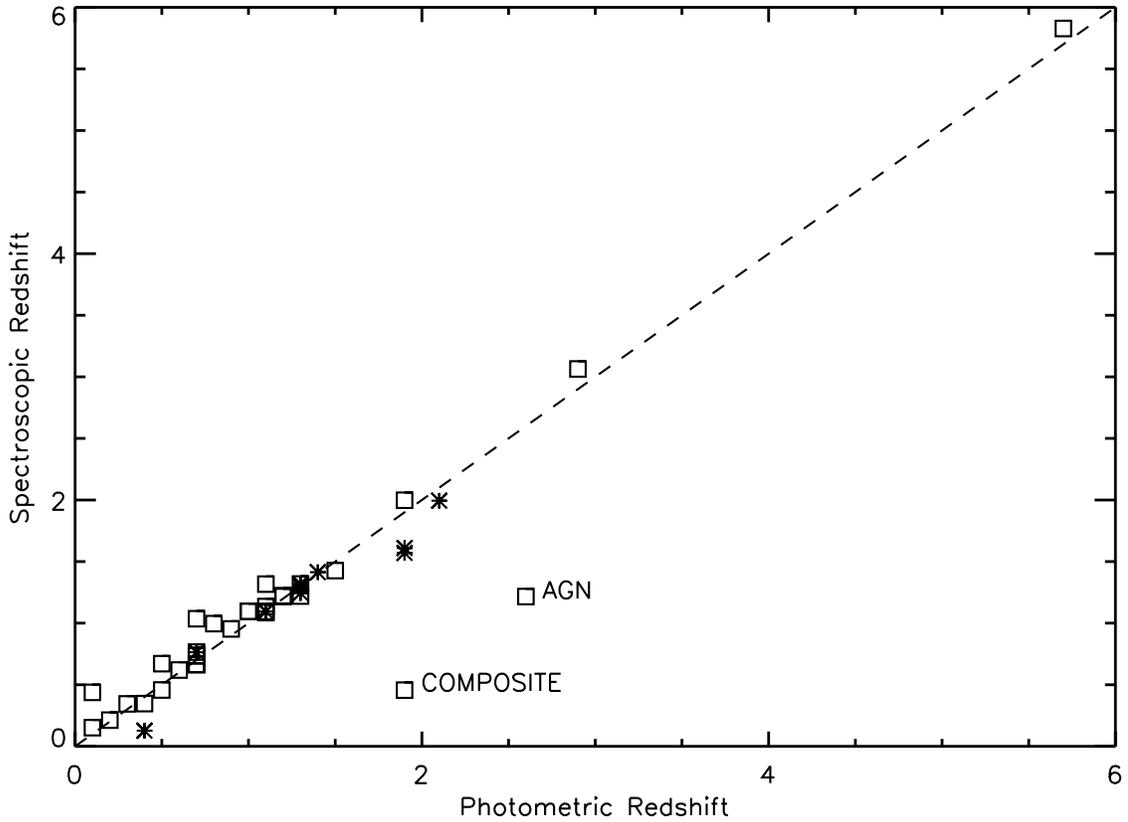}
\caption{A comparison of spectroscopic and photometric redshifts determined
in this study. The spectroscopic redshifts come from published an unpublished 
redshifts from a list 
utilized in the GOODS program. The square symbols indicate spectroscopic
redshifts rated excellent and asterisks redshifts rated good.  The
dashed line has a slope of 1 and is not a fit to the data.  The galaxy marked
AGN is an Active Galactic Nucleus and the galaxy marked COMPOSITE is described
in the text.  Some symbols closely overlie each other \label{fig-zcmp}}
\end{figure}

An additional list of published and unpublished spectroscopic redshifts
was prepared as part of the GOODS program. As a check on the validity of the 
photometric redshifts we compared the photometric and spectroscopic redshifts for 
the galaxies that had spectroscopic redshifts that were listed as equivalent to good 
or excellent in the heterogeneous rating methods of the lists. This resulted in 
46 galaxies including, the VIMOS galaxies, 
and the comparison is shown in Figure~\ref{fig-zcmp}. The references for the published
redshifts are (\citet{LeF04},\citet{szo04},\citet{mig05},\citet{van05}). Only one galaxy
redshift came from an unpublished observation. For the most part there is good
correspondence between the spectroscopic and photometric redshifts with the
exception of two catastrophic failures marked AGN and COMPOSITE in Figure~\ref{fig-zcmp}.  
The AGN is the galaxy 2340 in our catalog and its optical image shows a clear
diffraction spike from the nucleus. There is no AGN SED in our templates that
can be used to get a good redshift.  The second galaxy marked COMPOSITE in the
figure is galaxy 2170 in our catalog. It is adjacent to galaxy 2218 $1.5\arcsec$ 
away which has a photometric redshift of 0.5 equivalent to the spectroscopic 
redshift of 0.456.  Although 2218 is 2 magnitudes fainter than 2170 it is possible
that some spectral features contaminated the ground based spectrum of 2170 which
had a slit width comparable to the separation of the galaxies.  Currently there
are not many spectroscopic redshifts in th NUDF for redshifts between 2.2 and 5.5.
The reader is referred to \citet{thm03} for a spectroscopic versus photometric
redshift comparison in the HDFN that has numerous examples in this region.  The
analysis method here is exactly the same except for the different ACS filters.  
The Lyman break enters the ACS F435W filter at a redshift of 2.1, therefore, for
most of the range where we do not have spectroscopic confirmation the primary
signature of photometric redshifts is well covered.

\begin{figure}
\plotone{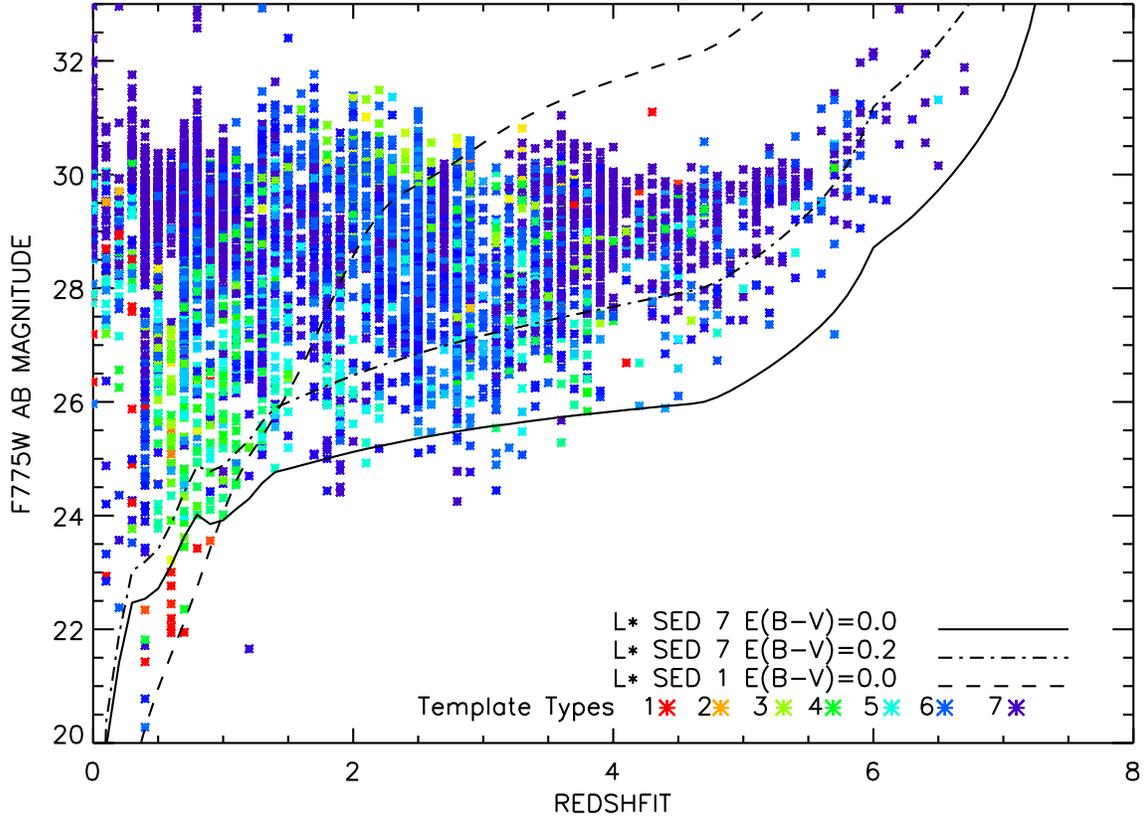}
\caption{Each galaxy in the source extraction is plotted in the F775W
versus redshift plane.  The symbols are color coded by SED template
type as shown in the symbol table at the bottom of the plot.  The 
density of the plot is such that the last galaxy plotted at a particular
location determines the color at that location.  The F775W AB
magnitude tracks for three different L$^*$ galaxies are shown on the plot
where L$^*$ is defined as L = $3.4 \times 10^{10}$ L$_{\sun}$.
\label{fig-mz}}
\end{figure}

Figure~\ref{fig-mz} shows a plot of the F775W AB magnitude versus redshift.
Superimposed on the figure are three lines showing the F775W magnitude of
an L$^*$ ($3.4 \times 10^{10}$ L$_{\sun}$ bolometric luminosity) galaxy at 
the redshifts in the plot for the highest temperature SED (7) with no
extinction, SED (7) with E(B-V) = 0.2 and the earliest SED (1) with
no extinction.  It is clear that vast majority of sources lie within the 
boundaries of these lines. In the previous papers (eg. \citet{thm03}) the
F160W magnitude was plotted instead of an optical magnitude.  The much 
higher signal to noise of the optical data in this case makes it more
instructive to use an optical magnitude. If we used the F160W band
more than 2/3 of the points would be null detections.

\subsection{Source 1810}

One source, ID = 1810, had photometric parameters of z = 2.7, E(B-V) = 1.0 and 
SED = 6.7, one of the hottest SEDs.  With these parameters it had an SFR of
27,463 M$_\sun$ per year, making the validity of these parameters very doubtful.
The fit had gone to the extremes of the extinction which predicted a very large
mid-infrared flux.  If the extinction is set to zero the appropriate fit parameters
are z = 1.2 and an SED = 1.1, a very early galaxy. Spitzer observations obtained in
the GOODS program \citep{dic06} showed IRAC
fluxes consistent with the early galaxy at redshift 1.2 interpretation and inconsistent
with  strong thermal dust emission at a redshift of 2.7.  A $3\sigma$ upper limit
of 18 microJanskys at 24 microns was provided by the MIPs observations.  For these
reasons the parameters for Source 1810 were set to their zero extinction values. With
these parameters its contribution to the SFR at redshift 1 is negligible.
No other source exhibited such an extreme SFR.

\subsubsubsection{The Nature of the Outliers}

The primary outlier in the redshift-magnitude diagram, Figure~\ref{fig-mz} is
object 2340 adjusted to its spectroscopic redshift, the AGN galaxy discussed above.  
This is object KX10 in the table of K band selected
xray sources in the CDFS \citep{cro01} and is identified as an AGN based on
the MgII and [OII] spectral features.  In fact the ACS images display faint
diffraction spikes, presumably from the nucleus of the AGN.

There are several galaxies near redshift 3 that exceed the unextincted L$^*$
magnitude but that is not unexpected.  In fact the low number of greater than L$^*$ galaxies 
is surprising when compared to the HDFN where there are 27 galaxies at redshifts
between 4 and 6.5 with greater than L$^*$ unextincted magnitudes. Galaxy 2888 is 
the only galaxy at z above 4 that has a greater than L$^*$ unextincted magnitude.  
Its photometric redshift is 5.7 and a spectroscopic redshift of 5.83, a SED 
template number of 5.9 and an E(B-V) value of 0.02.  This galaxy was known before
the position of the HUDF was picked and part of the decision on the location of 
the HUDF was the desire to include this galaxy. Detailed discussion of the 
scientific significance of the low number of high redshift greater than L$^*$
galaxies is beyond the scope of this paper other than to comment that if highly 
luminous galaxies are associated with very over dense regions then they may be more 
strongly correlated than lower luminosity galaxies.  

\subsection{Source Catalog}

A catalog of source properties is available in the electronic version of this paper.
The catalog is far to large to be published in the print version. Each row has 69 data
entries so it is not possible to print a stem of the catalog here.  The catalog is a
Right Ascension sorted catalog.  Along with its internal id number each source has
a cross reference to the Version 2 NICMOS UDF catalog available from MAST at the Space
Telescope Science Institute and the ACS UDF Version 1 catalog available from the same
source.  A 0 in the cross-reference indicates that there was no match.  Since the sources
were selected from a combination of the ACS and NICMOS fluxes, most of the cross references
to the Treasury catalog are 0.  For the ACS matching sources, the ACS catalog position was
required to be within 0.3 arc seconds of the NICMOS plus ACS position.  In some cases a 
different morphology in the infrared images may produce a false no match indication.
The $0.54\arcsec$ diameter aperture SFR is also provided.  This is the SFR from the flux in the
smallest photometric aperture.  This is not the flux used in the calculation of the SFR
in the previous section where the SFR for every detected pixel was used.  The sum of the
catalog SFRs will be substantially less than the SFR calculated in this paper.  Users can
adjust the rate by the ratio of the $0.54\arcsec$ diameter aperture fluxes to other fluxes 
of their choosing. The content and detailed format of the catalog is supplied in the 
electronic submission. All source numbers used in this paper refer to this catalog.

\section{Calculating the Star Formation Rates}

The procedure for calculating the SFRs is identical to the method
used in previous publications (\citet{thm01},\citet{thm03}) modulo
the changes described above in the extinction law, the template
SEDs and the details of the handling of the known redshift galaxies.  
For this reason only brief descriptions of the method are
given in this publication and the reader is referred to the publications
cited above for more detailed descriptions.

\subsection{The Star Formation Intensity Distribution} \label{ss-hx}

The star formation distribution function $h(x)$, first developed by
\citet{lan99}, and further discussed in \citet{lan02} has proved to be 
an excellent measure of the total star
formation rate at a given redshift range.  The distribution has been 
discussed several times but a brief description is given here as the
form of the distribution is not immediately intuitive.  The star formation
intensity $x$ is the SFR in solar masses per year per \emph{proper} square
kiloparsec.  This is calculated for each pixel that is considered part
of a galaxy.  These are the pixels marked in the detection image returned
by SE.  For a given redshift interval the distribution function $h(x)$
is the sum of all the proper areas in an interval of x, divided by that
interval and divided by the \emph{comoving} volume in cubic megaparsecs defined
by the boundaries of the observed field and the redshift interval \citep{lan99}.
The relation between the SFR and $h(x)$ is given by equation~\ref{eq-hx}

\begin{equation}
SFR = \int_{0}^{\infty}xh(x)dx
\label{eq-hx}
\end{equation}

\noindent which shows that the star formation rate is first moment of $h(x)$.
If the shape of $h(x)$ is invariant with redshift, the $h(x)$ as determined 
from low redshift observations can be used to correct the high redshift rates
for the losses due to surface brightness dimming.  This is done by sliding a
smoothed version of the distribution as measured in the redshift 1 bin
vertically until it matches the bright, large $x$ end of the distribution and
then performing the integral in equation~\ref{eq-hx} over the adjusted $h(x)$.
The smoothing is a 3 point smoothing for $\log h(x) \geq 10^{-4}$ which
corresponds to $\log x \leq 1$ in the redshift 1 distribution.
The height of the vertical adjustment was calculated by requiring the average
of the smoothed $h(x)$ distribution at the 6 highest $\log x$ values for 
$\log x \leq 1$ to match the average of the of the observed $h(x)$ at the same 
6 $\log x$ values.  If there are less than 6 valid values in the observed $h(x)$ 
then only the valid points are used. A similar method is used for the distribution 
without extinction correction but with the smoothing done over higher $\log h(x)$ 
values and lower $\log x$ values with only the first point used in determining 
the vertical scaling. The validity of the invariance with redshift and the empirical 
reasons why this might be so are discussed in detail in \citet{thm02}. Briefly
the change in the SFR intensity brought on by the expected smaller size in 
galaxies is offset by the expected lowering of the value of M$_*$ in the Schecter
distribution of masses.  Also by using the extinction corrected SFR intensities
the corrections are far less than matching to the SFR intensities that have not
be corrected for extinction as was done in \citet{lan02}.

\begin{figure}
\epsscale{.8}
\plotone{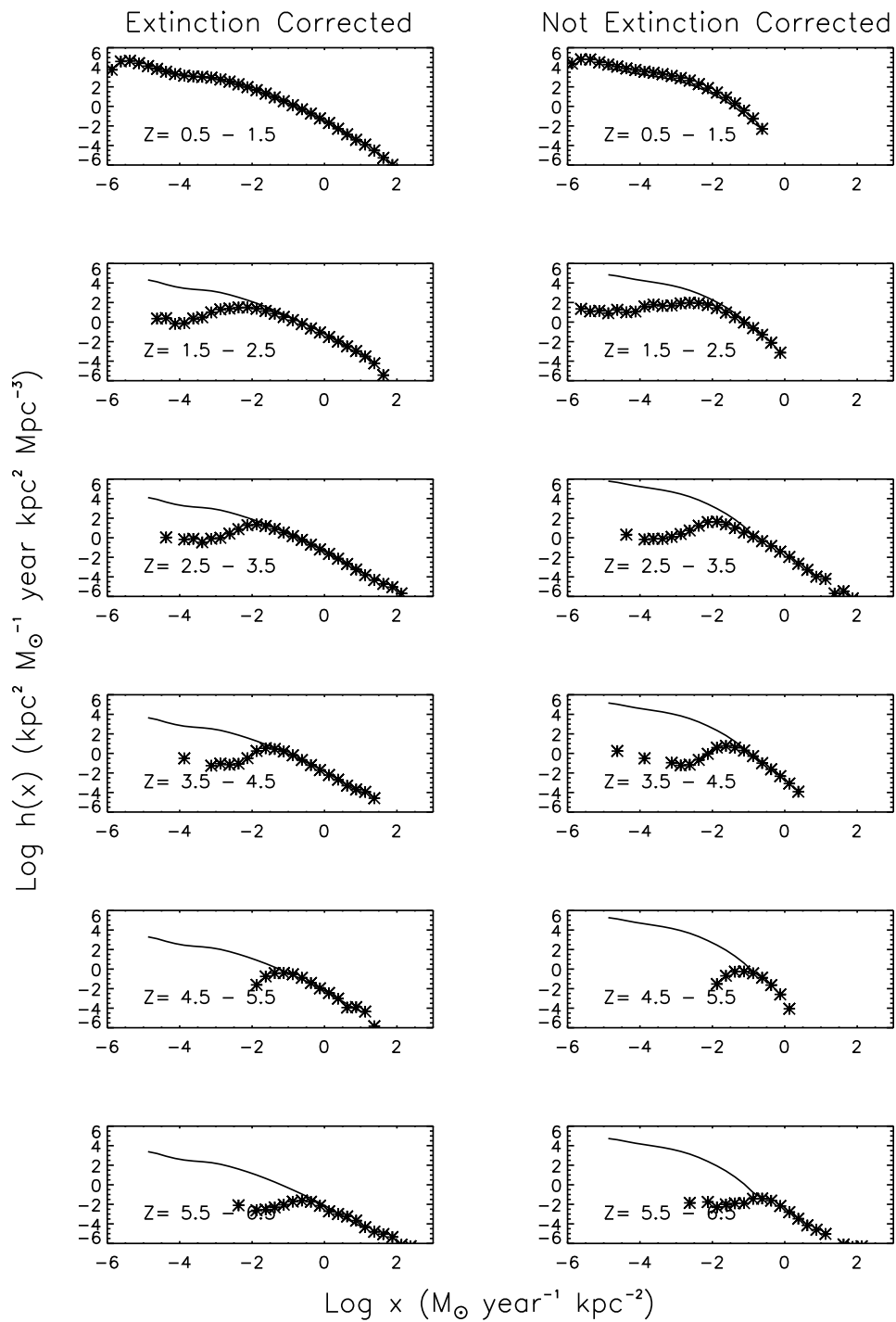}
\epsscale{1}
\caption{Plots of the star formation intensity distribution for each
redshift bin.  The left hand column intensities have been corrected 
for extinction while the right hand column intensities have not to 
show the difference between corrected and non-corrected distributions.
\label{fig-hx}}
\end{figure}

\begin{figure}
\plotone{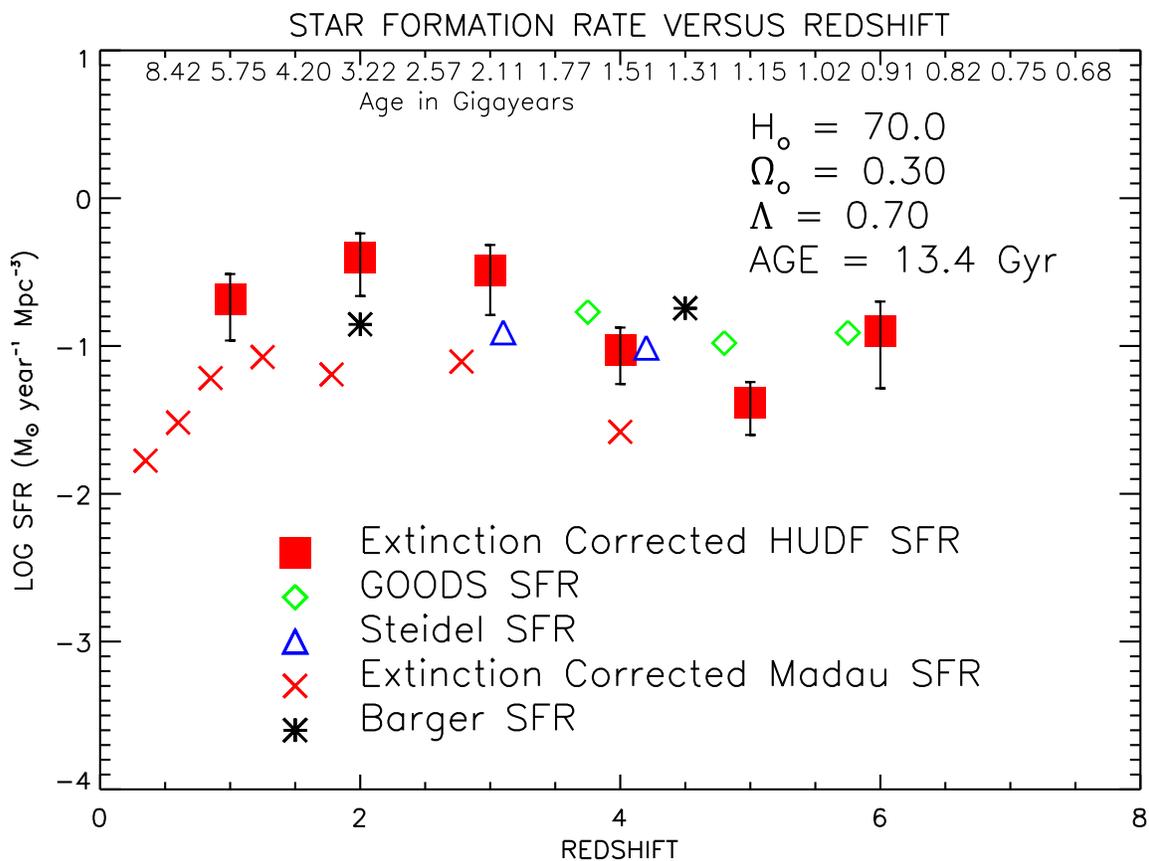}
\caption{The extinction corrected star formation history in the NICMOS 
region of the HUDF referred to as the NUDF in the text. The
SFR history for the much larger GOODS field \citep{gia04} is also shown
along with the original Madau data \citep{mad98}, the \citet{ste99} z of
3 and 4 ground based results and the \citet{bar00} sub-mm results. The
error bars are $1\sigma$.
\label{fig-sfr}}
\end{figure}

Figure~\ref{fig-hx} shows the measured SFR intensity distributions for redshifts
1 through 6 in unit redshift bins centered on the integer redshifts.  The left
hand column contains the distributions where all of the pixels have been corrected
for the extinction found for the galaxy they reside in and the right hand column
contains the distributions where no correction for extinction has been made. The
smoothed $h(x)$ is represented by the solid line and observed values are represented
by asterisks.  It is easy to see the effects of surface brightness dimming at redshifts
above 1 where the observed points dip below the smoothed line at low values of $\log x$. 
We will only use the results of the corrected distribution but show the uncorrected
rates for comparison.  Some of these comparisons are discussed in \S~\ref{ss-hxd}.
The individual pixel areas are roughly 1 square kiloparsec at redshifts of 1 and 
greater.  This is an optimal size as it represents roughly the smallest area over which
the Schmidt law is valid \citep{ken98}, which is one of the empirical laws the invariance
of $h(x)$ is based on \citep{thm02}.  The corrections on the star formation rate for
both extinction and surface brightness dimming by the $h(x)$ method are shown in 
Table~\ref{tab-sfr}.  The larger surface brightness dimming correction for redshift
2 may be due to a slight mismatch in the $h(x)$ fitting which can be seen for $\log x$ 
values less than 0 in Figure~\ref{fig-hx}.  The correction is less than the error
bars so there has been no attempt to adjust the value.

\subsection{The Star Formation Rates}

\begin{figure}
\plotone{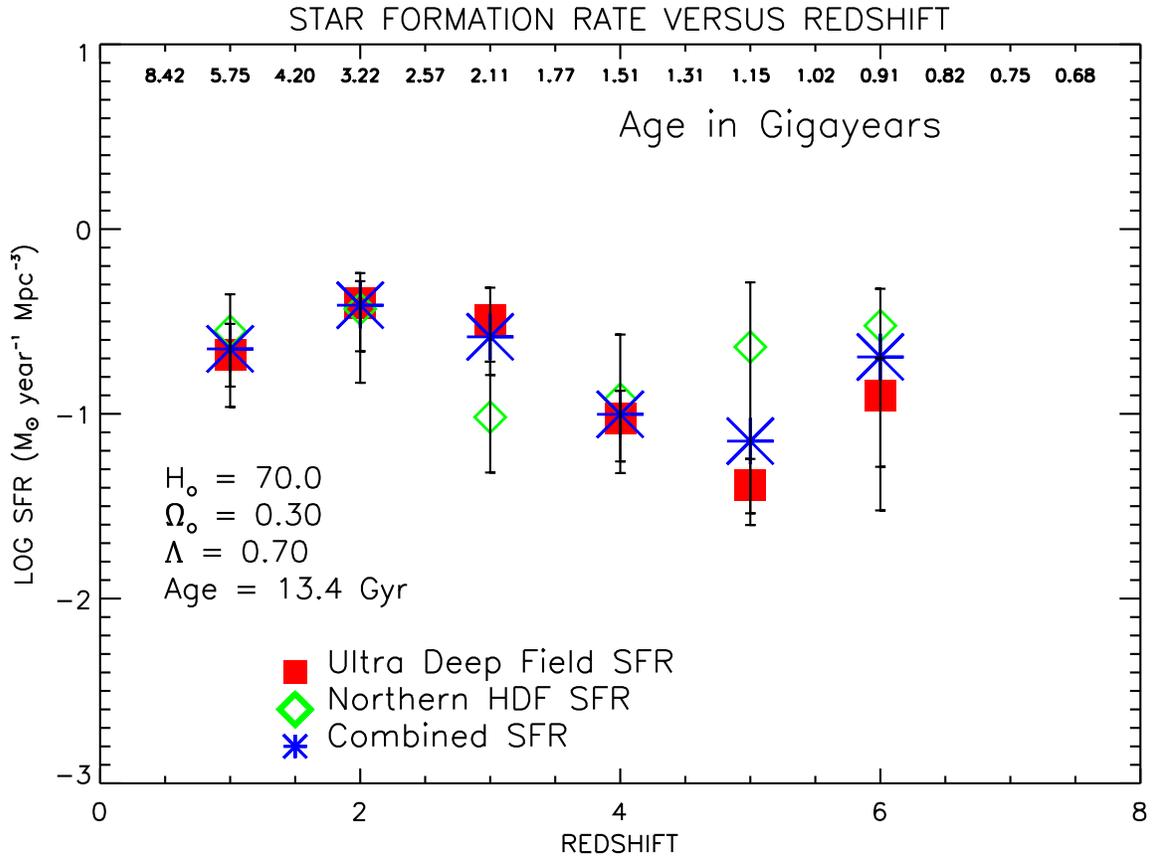}
\caption{The extinction corrected star formation history in the NICMOS region 
of the HUDF compared to the extinction corrected star formation history in the 
HDFN.  The HDFN SFRs are from \citet{thm03}. The
error bars are $1\sigma$. Note that the vertical scale is different from
Figure~\ref{fig-sfr}.
\label{fig-cmp}}
\end{figure}

Figure~\ref{fig-sfr} shows the star formation rates for the NICMOS region of the 
UDF binned in unit redshift bins centered on integer redshifts.  The redshift range
between 0.0 and 0.5 is not considered since small errors in redshift create very
large errors in the calculated SFR for objects that close.  The 1 $\sigma$ error bars 
are the rms sum of photometric, number and large scale structure errors as described 
in \S~\ref{ss-er}.  The rates are calculated as described in \S~\ref{ss-hx} with the
cosmological parameters H$_0$ = 70., $\Omega_0$ = 0.3 and $\Lambda$ = 0.7. All SFRs
from other works shown in the figure have been adjusted to this cosmology.  The 
main features of the star formation history in the NUDF is a slightly 
elevated SFR at redshifts 1-3 and a slightly depressed rate at redshifts 4-6.  The 
comparison with the star formation history in the HDFN is shown in Figure~\ref{fig-cmp}.
Although the histories appear to be somewhat dissimilar they are within 1 $\sigma$ of 
the HDFN error bars for all redshifts.  The smaller error bars in the UDF are due to 
the increased number of sources and the higher signal to noise of the ACS photometry.

\subsubsection{Errors in the SFR} \label{ss-er}

There are several sources of errors in determining the SFR history of the NUDF.  Some
of these can be quantitatively measured while others, such as extinction not conforming
to the extinction formula used in the analysis, are more difficult to measure.  The first
source of error addressed in photometric error.  Photometric error is addressed by
randomly altering the fluxes in each band for each source in a Gaussian distribution 
based on the measured 1$\sigma$ flux error in each band for the source and recomputing 
the SFR for each redshift bin.  This is done 100 
times using the 1$\sigma$ flux error computed by SE for each band.  This procedure 
incorporates not only the flux errors in the computed $1500$ Angstrom flux but also the changes
in redshift, extinction and derived SED that flux errors can introduce.  The SFR 
calculation is done in exactly the same manner as described in \S~\ref{ss-hx} which
gives the error in SFR for a given redshift bin rather than on a galaxy by galaxy
basis.  The photometric error in each redshift bin is calculated from the standard 
deviation of the 100 different SFRs for each bin. The photometric error is then root 
mean square combined with the shot noise error taken as

\begin{equation}
SFR_{shot noise} = \frac{\sqrt{\sum{sfr^2}}}{\sum{sfr}}
\label{eq-sn}
\end{equation}

The other source of error is large scale structure as described in \S~\ref{sss-ls}.
The total error, as shown in Figure~\ref{fig-sfr} is the root mean square of the 
photometric and number errors with the large scale structure error.  The contributions 
from each source of error for each redshift bin is tabulated in Table~\ref{tab-err}.

\subsubsection{Calculation of Large Scale Structure Error} \label{sss-ls}

The uncertainty in the star formation rate due to sample variance, also referred to as
large-scale structure, can be estimated by computing the variance in
the density field for the sample volume.  The variance of the density
field is given exactly by integrals of the two-point correlations;
higher-order correlations and non-Gaussianity enter only into higher
moments of the distribution of densities.

The survey volume in the case of the NUDF is very extended along the
line-of-sight compared to the transverse size and hence to the wavelength
of the fluctuations that dominate the variance.  This makes Limber's
equation a good approximation \citep{Lim53}
The angular power spectrum of a given
selection of objects is given in flat cosmologies by
\begin{equation}
P_2(K) = {1\over K} \int dk\;P(k) f(K/k) = 
\int {dr_a\over r_a^2} P(K/r_a) f(r_a)
\end{equation}
where $K$ is the angular wavenumber in rad$^{-1}$,
$r$ is a separation on the sky in $h^{-1}$~Mpc,
$k$ is the spatial wavenumber in $h$~Mpc$^{-1}$,
$P_2(K)$ is the angular power spectrum, and
$P(k)$ is the non-linear spatial power spectrum;
see \citet{Bau94} for more discussion and notation.
We have suppressed the redshift dependence of the 
power spectrum.
Cosmology enters through the coordinate distance
$r_a$ to a redshift $z$, and $r_a = K/k$.
Finally, $f(r)$ is the square of the probability distribution of 
the coordinate distance to the selected objects; this is trivially
related to the redshift distribution of the objects.  For 
objects spread uniformly in $r_a$ between two limits $r_1$ and $r_2$,
we simply have $f(r) = (r_2-r_1)^{-2}$ in that region and zero elsewhere.

For photometric-redshift selected samples at $z\gtrsim1$, it is usually the 
case that one has selected a sample for which the coordinate distance does not 
vary much through the slab.  In this case, one can treat $P$ and $r_a$ as
constant in the region where $f$ is non-zero.  This yields the form
\begin{equation}
P_2(K) = 
{P(K/r_a) \over r_a^2} \int dr_a f(r_a)
\approx
{P(K/r_a) \over r_a^2 \Delta r} 
\end{equation}
where the last form assumes a uniform distribution of a thickness
$\Delta_r = r_2-r_1$, which in turn is $c\Delta z/H(z)$.
It should be noted that distributions with softer cutoffs actually
yield smaller clustering amplitudes (but correlate neighboring slabs).

With the angular power spectrum of the selected objects in hand, one
can compute the variance in a given angular survey region.  This is 
easiest for a circular patch, for which the answer is simply
\begin{equation}
\sigma^2 = 
	{2\over \pi} \int K\;dK\; P_2(K) 
	    \left[J_1^2(Ka)\over Ka\right]^2
\approx	{2\over \pi \Delta r} \int k\;dk\; P(k) 
	    \left[J_1^2(kR)\over kR\right]^2
\end{equation}
where $a$ is the angular radius of the patch and
$R$ is the transverse radius of the patch (in length units, not angles).
The second equality uses the uniform approximation given above.
The difference between the variance of a square region and a circular one of the
same size is small, negligible for this estimate.  We therefore
treat the NUDF as a circle of $81''$ radius.

Using the standard $\Lambda$CDM cosmology and a simple model for the
non-linear power spectrum, we estimate that the rms variations in the
density for patches the size of the NUDF and unit redshift thickness
are about $0.3\sigma_8$ for $z=1$ to $z=4$, growing to about $0.4\sigma_8$
at $z=6$.  Here, $\sigma_8$ is the rms fluctuations of the galaxies in 
spheres of $8h^{-1}$~Mpc comoving radius.  Typically, one would 
estimate $\sigma_8\approx1$ for most populations of star-forming 
galaxies, although this may be a slight overestimate, as recent
clustering results at $z\gtrsim1$ have been tending to come in 
slightly smaller.  For example, the correlation lengths of order
$3-4h^{-1}$~Mpc in \citet{Coi04} and \citet{LeF05} would correspond
to $\sigma_8\approx0.8$.  However, the results of \citet{Ade05}
are $4-4.5h^{-1}$~Mpc, and so we keep $\sigma_8=1$ to be conservative.
In detail, different populations of galaxies 
have different clustering, and we are interested in the 
star-formation-weighted bias.
The systematic errors in Limber's equation and our 
subsequent approximations are small compared to the uncertainties in the
clustering amplitude.

Hence, we expect rms fractional uncertainties in the star formation rate
of about 30\% per unit redshift.  This is large enough that one does
not expect the distribution to be Gaussian---density fluctuations are
typically skew-positive---but our computation from the power spectrum
does get a correct measurement of the variance of the distribution. 
The results of the calculation for the area of the NUDF are given in 
Table~\ref{tab-err} in the column labeled LSS err. The error bars in 
Figure~\ref{fig-sfr} indicate the combined photometric and large scale 
structure errors.

\subsection{Galaxies that Contribute the Majority of the Star Formation} \label{ss-msf}

Table~\ref{tab-msfr} lists the number of galaxies that contribute $90\%$ of the 
SFR in each redshift bin and the total number of galaxies in the redshift bin. In all 
cases less than half of the galaxies contribute $90\%$ of the SFR. In most cases
less than $25\%$ of the galaxies contribute $90\%$ of the SFR.  It should be
emphasized that the calculation of galaxy SFRs was done by integrating the SFR
in all pixels defined by SE as part of a galaxy.  The total SFR at each redshift
was calculated by the method described in \S~\ref{ss-hx} which accounts for flux
missed in the source extraction by SE.

Comparison of the number of galaxies contributing 90\% of the SFR to the number of 
galaxies in Table~\ref{tab-msfr} indicates that the majority of star formation appears
to occur during periods of enhanced star formation in galaxies rather than constant
sustained star formation over a long period of time.  If the hierarchical model of
galaxy formation is invoked this may be during the time when mergers occur and leads
to the interesting speculation that perhaps the star formation history of galaxies
is also a monitor of the hierarchical formation history.  In that case the numbers 
in Table~\ref{tab-msfr} indicate the percentage of galaxies with magnitudes brighter
than the magnitude limit of the sample that are in the starburst/merger mode at that epoch.

There is clearly a deficit of high SFR galaxies at redshifts of 4 and 5 which is 
reflected in the lower total SRF at those redshifts.  The highest SFR galaxies
at redshifts of 4 and 5 have SFRs that are roughly 10 times less than the
SFRs for galaxies at lower redshifts.  The lack of high SFR regions is also
demonstrated in the $h(x)$ plots in Figure~\ref{fig-hx}.  As pointed out earlier this
is in contrast to the results found in the HDFN and may be the result of the 
effects of LSS.  Again this may reflect the conclusion that
the majority of total star formation occurs in the few galaxies with the highest
SFR and that the lack of such galaxies in a region of space and redshift range 
results in a lower total star formation for that region.  It may also be true
that if the star formation rate reflects the merger rate then the high SFR galaxies
may reside in denser regions of space and they may be more clustered than the
total distribution of matter in the universe.

\subsection{Discussion of Physical Insights from the $h(x)$ Distribution} \label{ss-hxd}

The star formation intensity distributions presented in Figure~\ref{fig-hx}
present a different way of looking at the history of star formation than the 
galaxy by galaxy view.  The proper areas of the pixels at redshifts of 1 and
beyond are roughly 1 sq. kiloparsec which is the area where Schmidt's law
relating SFR to gas column density becomes valid \citep{ken98}.  The star formation
intensity distribution shows the distribution of the star formation rate per
unit area as opposed to star formation rate per galaxy.  For example it shows
that at redshifts of 4 and 5 there is a lack of intense star formation areas
as well as a lack of high SFR galaxies.  Invoking the Schmidt law this may
mean that there is a lack of galaxies with areas of high gas density at these
redshifts in the NUDF region.  Somewhat hidden in the log-log plot over
a large range of values is that the intensity distribution at $x$ values greater
than 1 M$_{\sun}$ per year per proper sq. kiloparsec is dropping sharply
\citep{thm02} and that the probability of finding areas with SFRs of 100
M$_{\sun}$ per year per proper sq. kiloparsec is quite rare.

At fainter intensities it is clear that surface brightness dimming is removing
galaxies and parts of galaxies from detection.  For photon counting instruments
such as ACS and NICMOS the surface brightness dimming goes as $(1 + z)^{-4}$
which is reflected in the figure. Note however that in the star formation density
distribution derived from the redshift 1 results in the HDFN that 60\% of the 
total star formation occurs in the distribution greater than log(x) = -.25 and
95\% from the distribution greater than log(x) = -1.25 \citep{thm01} therefore
we are still seeing most of the star formation even at higher redshift. This
adds to our earlier conclusions that most of the star formation occurs in
high SFR galaxies by saying that most of the star formation occurs in high
SFR areas.

\section{Combined Star Formation Rates From the HDFN and the NUDF} \label{s-comb}

One of the major goals of the NUDF observations was to measure the SFR history
in a region that was completely uncorrelated with the HDFN.  Figure~\ref{fig-cmp}
shows the SFR history of the HDFN as determined by \citet{thm03} along with the
SFR history of the NICMOS UDF determined in this work.  Although the areas of the
two fields are roughly similar the UDF images, due to the ACS observations, are
considerably deeper.  To get the combined SFR history of the two fields we weighted
the SFR in each field by the number of galaxies in each redshift bin.  The error bars
in the HDFN values are from \citet{thm03}.  The average SFR
of the combined fields at z = 1-3 is 0.29 M$_{\sun}$ yr$^{-1}$ while the average at
z = 4-6 is 1.2 M$_{\sun}$ yr$^{-1}$. The SFRs at all redshifts are within 3$\sigma$ of
the average over all redshifts. Deep measurements over a much larger area are needed to 
accurately determine SFR history of the universe as opposed to just in the two small
deep fields.

\section{Conclusions} \label{s-conc}

The NUDF provides an independent measure of the star formation
history at high redshifts.  Unlike the HDFN, the NUDF appears to
have a paucity of very luminous and high SFR galaxies past a redshift of approximately
3.  This lack is probably due to variances in population due to large scale structure.
The variances between the fields highlights large scale structure as a major error
in the determination of the SFR history of the universe.  The combined data of the HUDF
and the HDFN show a peak in the SFR at a redshift of 2 with a decline toward higher 
redshifts.  Deep optical and infrared images over a very large area will be needed to 
quantify the SFR history of the universe.  Progress on this quest can be made if the 
infrared channel of the WFC3 becomes operational on HST.  True progress may require a 
dedicated mission similar to some of the missions developed as part of the Origins Probe
mission studies commissioned by NASA.

\acknowledgments

We wish to than an anonymous referee for helpful comments that improved the paper.
We wish to thank Eros Vanzella for providing unpublished redshifts that
have been used for spectroscopic comparison to our photometric redshifts.
This article is based on data from observations with the 
NASA/ESA Hubble Space Telescope, obtained at the Space Telescope Science 
Institute, which is operated by the Association of Universities for Research 
in Astronomy under NASA contract NAS 5-26555.  RIT, DE, XF and GI are 
funded in part by NASA Grant HST-GO-09803.01-A-G from the Space Telescope 
Science Institute.

\clearpage

\begin{deluxetable}{ccccccc}
\tablecaption{Aperture corrections for the $0.54\arcsec$ diameter aperture. As
expected the aperture corrections increase starting with the F850LP band where
a combination of scattering and the diffraction width limited PSF start to become 
significant relative to the aperture diameter.  The NICMOS aperture corrections
were derived from stellar observations as described in the text.
\label{tab-ap}}
\tablewidth{0pt}
\scriptsize
\tablehead{\colhead{Filter} & \colhead{F445W} & \colhead{F606W} &
\colhead{F775W} & \colhead{F850LP} & \colhead{F110W} & \colhead{F160W}
}

\startdata
Correction Factor & 1.15 & 1.14 & 1.15 & 1.21 & 1.43 & 1.55 \\
\enddata
\end{deluxetable}

\clearpage

\begin{deluxetable}{ccccc}
\tablecaption{SFR corrections for extinction and surface brightness dimming.
The volume in the SFR units is the comoving volume. 
\label{tab-sfr}}
\tablewidth{0pt}
\scriptsize
\tablehead{\colhead{ } & \colhead{ } & \colhead{ } &
\colhead{Extinction and} & \colhead{Missing} \\
\colhead{} & \colhead{Uncorrected} & \colhead{Extinction Corrected} & 
\colhead{Dimming Corrected} &\colhead{Luminosity \tablenotemark{a}} \\
\colhead{z} & \colhead{SFR (M$_{\sun}$ $yr^{-1}$ $Mpc^{-3}$)} & 
\colhead{SFR (M$_{\sun}$ $yr^{-1}$ $Mpc^{-3}$)} & 
\colhead{SFR (M$_{\sun}$ $yr^{-1}$ $Mpc^{-3}$)} & \colhead{($\%$)}
}

\startdata
1 & 0.017 & 0.21 & 0.21 & 0 \\
2 & 0.021 & 0.30 & 0.40 & 24 \\
3 & 0.13 & 0.29 & 0.32 & 10 \\
4 & 0.027 & 0.082 & 0.094 & 13 \\
5 & 0.014 & 0.034 & 0.041 & 17 \\
6 & 0.019 & 0.10 & 0.13 & 17 \\
\enddata
\tablenotetext{a}{This is the percentage of missed luminosity between 
the extinction corrected SFR and the extinction corrected SFR which has 
also been corrected for surface brightness dimming.}
\end{deluxetable}
\clearpage

\begin{deluxetable}{cccccc}
\tablecaption{Contributions to the total SFR error in M$_{\sun}$ yr$^{-1}$ 
Mpc$^{-3}$. The column labeled Std. Dev. is the standard deviation of the 
SFRs computed in the 100 calculations of the SFR with random photometric
errors added to the observed fluxes. The column marked SFR shot noise is the
error computed in equation~\ref{eq-sn}. The total errors are the root mean 
square sum of the individual errors.  All errors are in units of  M$_{\sun}$ yr$^{-1}$ 
Mpc$^{-3}$ not fractions which accounts for the lower errors in regions of low
SFR.
\label{tab-err}}
\tablewidth{0pt}
\scriptsize
\tablehead{\colhead{Z} & \colhead{SFR} & \colhead{Std. Dev.} & 
\colhead{SFR shot noise} & \colhead{LSS err.} & \colhead{Total error.}
}
\startdata
1 & 0.208 & 0.036 & 0.056 & 0.073  & 0.099 \\
2 & 0.398 & 0.064 & 0.100 & 0.140  & 0.180 \\
3 & 0.322 & 0.057 & 0.110 & 0.110  & 0.160 \\
4 & 0.094 & 0.011 & 0.017 & 0.033  & 0.039 \\
5 & 0.041 & 0.006 & 0.006 & 0.014  & 0.016 \\
6 & 0.126 & 0.026 & 0.053 & 0.044  & 0.074 \\
\enddata
\end{deluxetable}
\clearpage

\begin{deluxetable}{ccccccc}
\tablecaption{The table lists the number of galaxies that contribute $90\%$
of the total SFR for a redshift bin and the total number of galaxies in the
redshift bin. The SFR for the galaxy with the highest, third highest, and
tenth highest SFR along with the SFR of the galaxy with the lowest SFR, but 
still in the group of galaxies that contribute $90\%$ of the SFR, is also given.
\label{tab-msfr}}
\tablewidth{0pt}
\scriptsize
\tablehead{\colhead{ } & \colhead{z=1} & \colhead{z=2} & \colhead{z=3} & \colhead{z=4}
& \colhead{z=5} & \colhead{z=6}
}
\startdata
Galaxies contributing $90\%$ of the SFR & 182 & 118 & 171 & 163 & 98 & 14 \\
Total number of galaxies & 1518 & 929 & 928 & 509 & 218 & 64 \\
Highest SFR in M$_{\sun}$ yr$^{-1}$ & 425 & 899 & 1600 & 138 & 43 & 309 \\
Third Highest SFR & 260 & 582 & 347 & 81 & 20 & 59 \\
Tenth Highest SFR & 33 & 96 & 56 & 22 & 11 & 21 \\
Lowest SFR in the $90\%$ group & 1.3 & 4.1 & 2.9 & 1.1 & 0.8 & 12 \\
\enddata
\end{deluxetable}
\clearpage

\clearpage

\clearpage

\clearpage

\clearpage

\clearpage

\clearpage

\clearpage

\clearpage

\begin{thebibliography}{}

\bibitem[Adelberger et al.(2005)]{Ade05}
    Adelberger, K.L., Steidel, C.C., Pettini, M., Shapley, A.E., 
    Reddy, N.A., \& Erb, D.K., 2005, \apj, 619, 697

\bibitem[Aussel et al.(1999)]{aus99} Aussel, H. et al. 1999, \aap, 342, 313

\bibitem[Barger, Cowie \& Richards(2000)]{bar00} Barger, A.J., Cowie, L.L.,
	Richards, E.A. 2000, \aj, 119, 2092

\bibitem[Baugh \& Efstathiou(1994)]{Bau94}
    Baugh, C.M., \& Efstathiou, G. 1994, \mnras, 267, 323

\bibitem[Beckwith et al.(2006)]{bec06} Beckwith, S. et al. 2005, in preparation

\bibitem[Bertin \& Arnouts(1996)]{ber96} Bertin,E., \& Arnouts, S.
	1996, \aap, 117, 393

\bibitem[Bouwens et al.(2004)]{bou04} Bouwens, R.J. et al. 2004, \apjl, 616, L79

\bibitem[Bouwnes et al.(2005)]{bou05} Bouwens, R.J. et al. 2004, \apj, in press, 
	astro-ph/0509641 v4

\bibitem[Bruzual \& Charlot(1996)]{bru96} Bruzual, G. \& Charlot, S., 1996

\bibitem[Bruzual \& Charlot(2003)]{bru03} Bruzual, G. \& Charlot, S., 2003,
	\mnras, 344, 1000

\bibitem[Calzetti, Kinney, \& Storchi-Bergmann(1994)]{cal94} Calzetti, D.,
	Kinney, A.L., \& Storchi-Bergmann, T. 1994, \apj, 429, 582

\bibitem[Calzetti et al.(2000)]{cal00} Calzetti, D., et al. 2000, \apj, 533, 682 

\bibitem[Chabrier(2003)]{cha03} Chabrier, G. 2003, \pasp, 115, 763

\bibitem[Coil et al.(2004)]{Coi04}
    Coil, A.L., et al., 2004, \apj, 609, 525

\bibitem[Coleman, Wu \& Weedman(1980)]{col80} Coleman, G. D., Wu, C. C., Weedman, D. W.
	1980, \apjs, 43, 393

\bibitem[Croom, Warren \& Glazebrook(2001)]{cro01} Croom, S.M., Warren, S.J. \&
	Glazebrook, K. 2001, \mnras, 328, 1506

\bibitem[Dickinson(2000)]{dic00} Dickinson, M. 2000, Phil. Trans. R. Soc. London A.,
	358, 2001

\bibitem[Dickinson et al.(2006)]{dic06} Dickinson,  M. et al. 2006, in preparation
\bibitem[Fernandez-Soto, Lanzetta \& Yahil(1999)]{fer99} Fernandez-Soto, A., Lanzetta,
	K.M., \& Yahil, A. 1999, \apj, 513, 34

\bibitem[Fruchter and Hook(2002)]{fru02} Fruchter, A.S. \& Hook, R.N. 2002, \pasp, 114, 144

\bibitem[Giavalisco et al.(2004)]{gia04} Giavalisco, M. et al. 2004, \apjl, 600, L103

\bibitem[Kennicutt(1998)]{ken98} Kennicutt, K.C., Jr. 1998, \araa, 36, 189

\bibitem[Lanzetta, Yahil \& Fernandez-Soto(1996)]{lan96} Lanzetta, K.M., Yahil,
	A., \& Fernandez-Soto, A. 1996, \nat, 381, 79

\bibitem[Lanzetta et al.(1999)]{lan99} Lanzetta, K.M. et al. 1999, in ASP Conf.
	Ser. 191, Photometric Redshifts and High-Redshift Galaxies, ed. R.J. 
	Weymann, L.J. Storrie-Lombardi, M. Sawicki, \& R.J. Brunner 
	(San Francisco ASP), 223

\bibitem[Lanzetta et al.(2002)]{lan02} Lanzetta, K.M. et al. 2002, \apj, 570, 492

\bibitem[Le F\`evre et al.(2004)]{LeF04}Le F\`evre, O., et al. 2004, \aap, 428, 1043 

\bibitem[Le F\`evre et al.(2005)]{LeF05}Le F\`evre, O., et al., 2005, \aap, 439, L877

\bibitem[Lilly et al.(1996)]{lil96} Lilly, S.J. et al. 1996, \apjl, 460, L1

\bibitem[Limber(1953)]{Lim53}
    Limber, D.N. 1953, \apj, 117, 134

\bibitem[Madau, et al.(1996)]{mad96} Madau, P. et al. 1996, \mnras, 283, 1388

\bibitem[Madau, Pozzetti, \& Dickinson(1998)]{mad98} Madau, P., Pozzetti, L., 
	\& Dickinson, M. 1998, \apj, 498, 106

\bibitem[Mignoli et al.(2005)]{mig05} Mignoli, M. et al. 2005, \aap, 437, 883

\bibitem[Persson et al.(1998)]{per98} Persson, S.E. et al. 1998, \aj, 116, 2475

\bibitem[Pirzkal et al.(2005)]{pir05} Pirzkal, N. et al. 2005, \apj, 622, 319

\bibitem[Sirianni et al.(2005)]{sir05} Sirianni, M. et al. 2005, \pasp, 117, 1049

\bibitem[Stanway, Bunker \& McMahon(2003)]{sta03} Stanway, E.R., Bunker, A.J.,
	McMahon, R.G. 2003, \mnras, 342, 439

\bibitem[Steidel et al.(1999)]{ste99} Steidel, C.C., Adelberger, K.L., 
	Giavalisco, M., Dickinson,M., \& Pettini, M. 1999, \apj, 519, 1

\bibitem[Stiavelli, Fall, \& Panagia(2004)]{sti04} Stiavelli, M., Fall, S.M.,
	\& Panagia, N. 2004, \apj, 600, 508

\bibitem[Szalay, Connolly \& Szokoly(1999)]{sza99} Szalay, A.S., 
	Connolly, A.J., \& Szokoly, G.P. 1999, \aj, 117, 68

\bibitem[Szokoly et al.(2004)]{szo04} Szokoly, G. P. et al. 2004, \apjs, 155, 271

\bibitem[Thompson et al.(1999)]{thm99} Thompson et al. 1999, \aj, 117, 17

\bibitem[Thompson et al.(2001)]{thm01} Thompson, R.I., Weymann, R.J. \& 
	Storrie-Lombardi, L.J. 2001, \apj, 546, 694

\bibitem[Thompson(2002)]{thm02} Thompson, R.I. 2002, \apjl, 581, L85

\bibitem[Thompson(2003)]{thm03} Thompson, R.I. 2003, \apj, 596, 748

\bibitem[Thompson et al.(2005)]{thm05} Thompson, R.I. et al. 2005, \aj, 130, 1

\bibitem[Thompson(2005b)]{thm05b} Thompson, R.I. 2005 in the Proceedings of the
	2005 HST Calibration Workshop, A. Koekemoer, P. Goudfrooij, and L. Dressel, eds., 
	Space Telescope Science Institute, in press.

\bibitem[Vanzella et al.(2005)]{van05} Vanzella, E. et al. 2005, \aap, 434, 53

\bibitem[Williams et al.(1996)]{wil96} Williams, R.E., et al. 1996, \aj, 112, 1335

\bibitem[Yan et al.(1999)]{yan99} Yan, L. et al. 1999, \apj, 519, L47

\bibitem[Yan \& Windhorst(2004)]{yan04} Yan, H. \& Windhorst, R.A. 2004, \apjl, 612, L93

\end{thebibliography}
\end{document}